\documentclass{cmbe24} 
\usepackage{amsmath,amsfonts,amssymb}
\usepackage{graphicx,wrapfig}
\usepackage[hyphens]{xurl}
\usepackage{hyperref}

\title{Pharmacometrics Modeling via Physics-Informed Neural Networks: Integrating Time-Variant Absorption Rates and Fractional Calculus for Enhancing Prediction Accuracy}

\author[1]{Nazanin Ahmadi}
\author[2,3]{Shupeng Wang}
\author[3]{George Karniadakis}
\affil[1]{Center for Biomedical Engineering, Brown University, Providence, RI, USA, \texttt{Nazanin@Brown.edu}}
\affil[2]{School of Mathematics, Shandong University, Jinan, China,
\texttt{201911826@mail.sdu.edu.cn}}
\affil[3]{Division of Applied Mathematics, Brown University, Providence, RI, USA, \texttt{George\_karniadakis@Brown.edu}}

\summary{We present a novel method to improve pharmacokinetics modeling, an essential step of drug development. Conventional models frequently fail to fully represent the intricacies of drug absorption and distribution, which limits their predictive abilities required for personalized treatment strategies. Our methodology introduces two innovations to enhance modeling accuracy: 1. Time-varying parameters: this approach is designed to accommodate the dynamic nature of drug absorption rates. 2. Fractional calculus in representing delayed drug response. This approach effectively captures anomalous diffusion phenomena, surpassing traditional models in describing drug delayed response without the need for extensive compartmentalization.}


\begin{document}

\section{Introduction}
Pharmacokinetic models are essential in drug discovery and clinical development, optimizing dose regimens and validating therapeutic responses. Traditional approaches typically utilize integer-order differential equations with constant parameters, which may not capture the memory effects prevalent in many drug delivery scenarios. In our study, we employ a three-compartment model for canines by Uno et al. \cite{Talaporfin}, which includes plasma, interstitial, and cellular compartments to detail drug dynamics.

Recent advancements in Physics-Informed Neural Networks (PINNs) \cite{PINNs} have revolutionized parameter estimation for systems exhibiting nonlinear dynamics, numerous undetermined parameters, and a scarcity of experimental data. The introduction of fractional PINNs (fPINNs) in \cite{fPINNs} extends these capabilities to fractional differential equations, enhancing modeling flexibility further. The AI-Aristotle framework \cite{AI-Aristotle} first integrated PINNs into pharmacokinetic modeling, setting a precedent for using these models to solve complex inverse problems. Our study builds on this foundation, using PINNs and fPINNs to improve accuracy and yield new insights in pharmacokinetic modeling.

\subsection{Three-Compartment Mathematical Modeling of Pharmacokinetics}

The system of Equation 1-3 introduces a three-compartment model for pharmacokinetics that includes compartments for plasma, interstitial space, and cells. Here, 0 is used to represent areas outside the model, while 1, 2, and 3 refer to the plasma, interstitial, and cell compartments, respectively. The model uses differential equations (1)--(3) to express the drug concentration within these compartments. For each compartment, represented by ``i", where i equals 1, 2, or 3, $V_i$ and $C_i$ stand for the volume and concentration of the compartment, respectively. The model also defines the rate of drug excretion from the plasma compartment as $k_{10}$, along with the transfer rates between compartments, denoted as $k_{12}$, $k_{21}$, $k_{23}$, and $k_{32}$.

\begin{align}
V_1 \frac{dC_1}{dt} &= -(k_{10} + k_{12})V_1C_1 + k_{21}V_2C_2, \\
V_2 \frac{dC_2}{dt} &= k_{12}V_1C_1 - (k_{21} + k_{23})V_2C_2 + k_{32}V_3C_3, \\
V_3 \frac{dC_3}{dt} &= k_{23}V_2C_2 - k_{32}V_3C_3.
\end{align}

\section{Methodology}
In this study, we aim to refine the estimation of parameters within a system governed by ordinary differential equations (ODEs). Traditionally, numerical solvers like the \texttt{fmincon} toolbox in MATLAB are utilized for this purpose. Our methodology differs by integrating two novel strategies to improve model accuracy: (i) the introduction of time-varying parameters to capture the dynamic nature of drug absorption rates, and (ii) the adoption of fractional calculus to better represent the lag in drug response. We explore the PINNs as an innovative solver. We compare the effectiveness of these methodologies, including PINNs and fPINNs, in accurately modeling the system of ODEs for Talaporfin sodium PK model.

The constants $[v_1, v_2, v_3]$ are maintained at $[394, 251, 970]$ mL, based on histological evaluations cited in \cite{TalaporfinInterstitial}, with supplementary data from \cite{Talaporfin}. Within the PINNs framework, the Fractional Finite Difference Method (FFDM) is applied to generate numerical solutions, particularly for two versions of Fractional Pharmacokinetics Models introduced in ~\cite{AmiodaroneDiffusion}.

\subsection{PINNs with Time-varying parameter}
Herein, we address the dynamic nature of drug absorption rates by treating \( k_{12} \), the primary rate constant, as a time-dependent function. This approach allows us to account for the time-delayed response of the drug. We tackl the inverse problem of simultaneously determining the values of all unknown system parameters, including \( k_{10} \), \( k_{21} \), \( k_{23} \), and \( k_{32} \), in conjunction with \( k_{12}(t) \).

\subsection{fPINNs}
fPINNs is a variant of PINNs used to solve PDEs or ODEs. We incorporate numerical differentiation formulas from fractional calculus for representing fractional operators, while leveraging automatic differentiation for integer-order operators. This hybrid approach introduces a composite effect where discretization, sampling, Neural Network approximation, and optimization errors collectively influence the convergence properties of fPINNs. The foundation of this approach lies in Caputo's definition of the fractional derivatives operator, denoted as \(^{C}D^{\alpha}\), where \(\alpha\) represents the fractional order. We infer the fractional order $\alpha$ along with the constant values of the unknown parameters $k_{10}$, $k_{12}$, $k_{21}$, $k_{23}$, and $k_{32}$.\\
\begin{itemize}
    \item[a)] The commensurate fractional three-compartmental pharmacokinetic (PK) model is described by the following set of fractional differential equations, where $\,^{C}_{0}D^{\alpha}_{t}$ denotes the Caputo fractional derivative of order $\alpha$:
    \begin{align}
    V_1\,^{C}_{0}D^{\alpha}_{t} C_1 &= -(k_{10} + k_{12})V_1C_1 + k_{21}V_2C_2, \\
    V_2\,^{C}_{0}D^{\alpha}_{t} C_2 &= k_{12}V_1C_1 - (k_{21} + k_{23})V_2C_2 + k_{32}V_3C_3, \\
    V_3\,^{C}_{0}D^{\alpha}_{t} C_3 &= k_{23}V_2C_2 - k_{32}V_3C_3.
    \end{align}
    \item[b)] The implicit non-commensurate fractional three-compartmental PK model introduces a different fractional order in one of the compartments. The model equations are written as:

    \begin{align}
    V_1 \frac{dC_1}{dt} &= -(k_{10} + k_{12})V_1 C_1 + k_{21}V_2 C_2, \\
    V_2 \frac{dC_2}{dt} &= k_{12}V_1 C_1 - (k_{21} + k_{23})V_2 C_2 + k_{32}V_3 \,^{C}_{0}D^{1-\alpha}_{t} C_3, \\
    V_3 \frac{dC_3}{dt} &= k_{23}V_2 C_2 - k_{32}V_3 \,^{C}_{0}D^{1-\alpha}_{t} C_3.
    \end{align}
\end{itemize}

\section{Results and conclusions}
We evaluated the effectiveness of photodynamic therapy in the myocardial interstitial space by developing a three-compartment pharmacokinetic model. This model aims to predict the interstitial concentration of talaporfin sodium in the canine, based on variations in its plasma, interstitial, and cellular levels. We integrate fractional calculus and PINNs with conventional pharmacokinetic modeling, employing three distinct approaches and the results for inferred values of parameters in each model are shown in Table~\ref{tab:parameters}.\\

\begin{table}[ht]
    \centering
    \begin{tabular}{l|c|c|c}
        \hline
        Parameter & PINNs (Mean \(\pm\) Std Dev) & fPINNs (a) & fPINNs (b)  \\
        \hline
        \(\alpha\) & - & 0.99168 & 0.67429 \\
        \(k_{10}\) & 2.332 \(\pm\) 0.019  & 2.43033 & 2.37014  \\
        \(k_{12}\) & Time variant& 4.34626 & 4.74927  \\
        \(k_{21}\) & 9.272 \(\pm\) 0.585  & 10.01725 & 10.85011 \\
        \(k_{23}\) & 7.941 \(\pm\) 0.070 & 3.42174 & 5.83796  \\
        \(k_{32}\) & 4.112 \(\pm\) 0.026 & 1.32857 & 3.44299  \\
        \hline
    \end{tabular}
    \caption{Comparison of parameters for fPINNs (a), fPINNs (b), and PINNs for 5 runs.}
    \label{tab:parameters}
\end{table}

\begin{figure}[h]
    \centering
    \includegraphics[width=0.6\textwidth]{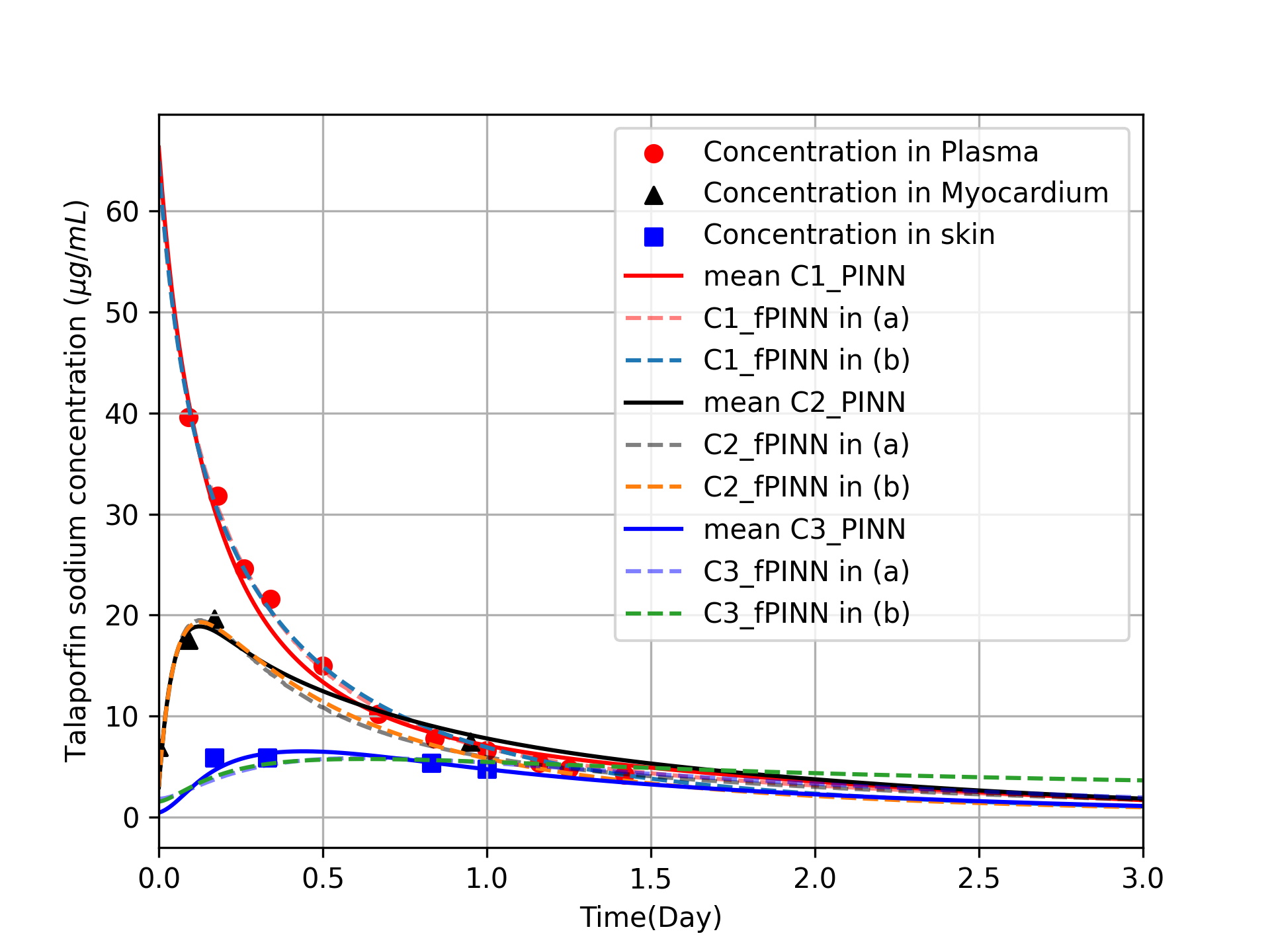}
    \caption{Comparison of the solutions to the system of ODEs using different approaches.}
    \label{fig:concentration}
\end{figure}

\begin{figure}[h]
    \centering
    \includegraphics[width=0.6\textwidth]{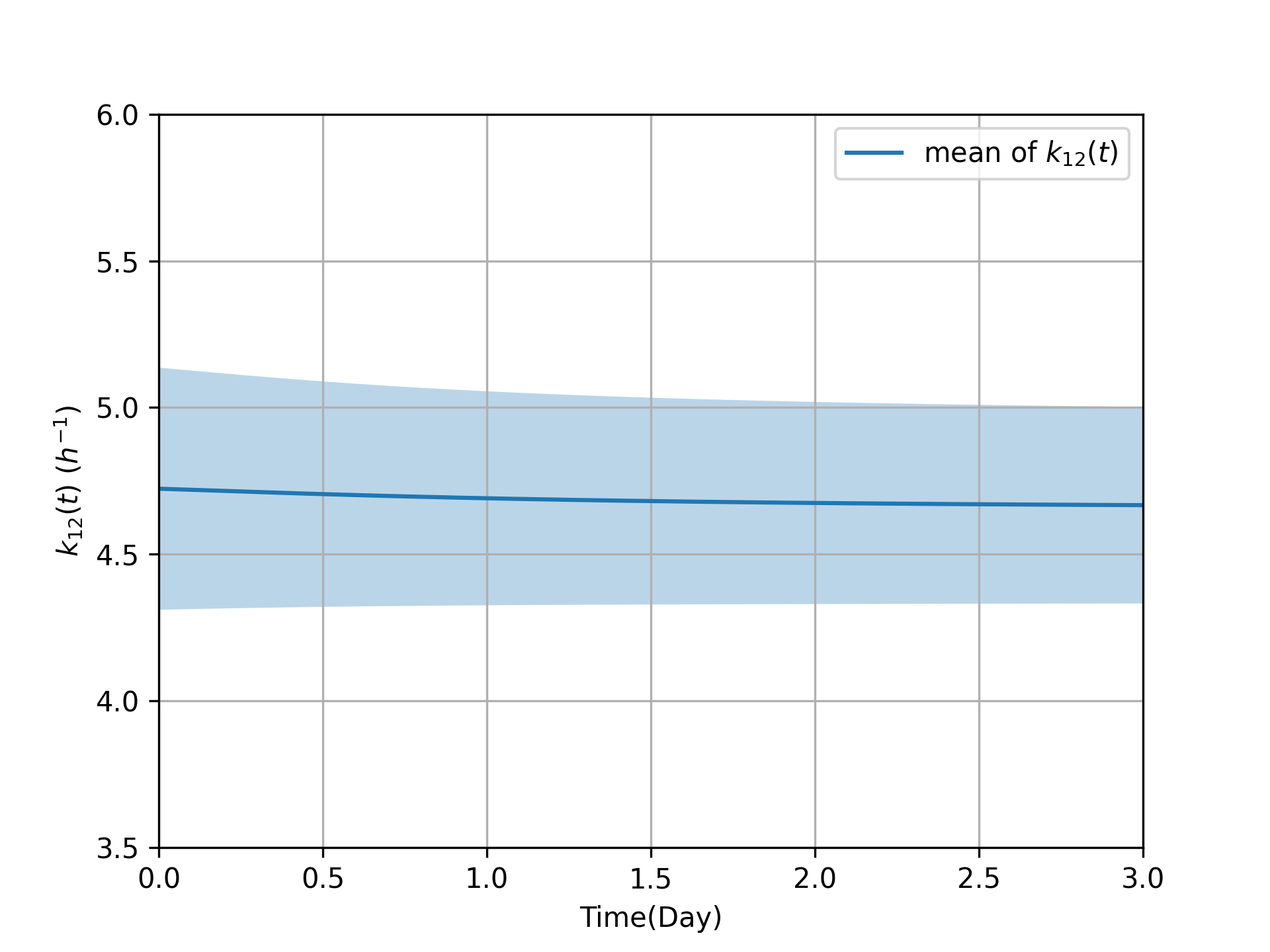}
    \caption{Variation of k12 as a function of time (mean and standard deviation.}
    \label{fig:k12}
\end{figure}

Figure~\ref{fig:concentration} shows the solution using different approaches, as well as the data points we had for each compartment. Meanwhile, Figure~\ref{fig:k12} illustrates the mean and standard deviation of $k_{12}$ as a function of time using the PINNs method. We can observe that the value of $k_{12}$ does not change significantly. However, even this minor variation is sufficient to represent the delay in the model.

We evaluated the fitting quality of three approaches using the $R^2$ score and compared the Mean Absolute Error (MAE) values for three concentrations as shown in Table~\ref{tab:mae_values}. PINNs showed lower MAE for concentrations C2 and C3, while fPINNs yielded better results for C1. The final R2 scores were similar across all methods. Further systematic studies are required to explain this observed sensivity in solution approximation between the fractional- and integer-order model.


\begin{table}[h]
\centering
\begin{tabular}{l|c|c|c|c}
\hline
Approach & MAE (C1) & MAE (C2) & MAE (C3) & $R^2$ Score \\
\hline
PINNs & 1.281 & 1.695 & 0.438 & 0.98\\
fPINNs (a) & 0.640 & 1.872 & 0.987& 0.98  \\
fPINNs (b) & 0.604 & 1.795 & 0.896& 0.98\\
\hline
\end{tabular}
\caption{The comparison of Mean Absolute Error (MAE) values for each approach and different compartments.}
\label{tab:mae_values}
\end{table}

A significant contribution of our work is the introduction of fPINNs to pharmacokinetics, marking its inaugural application in this field. By leveraging neural networks, we automated the optimization of parameters, including fractional orders, during training. This streamlined the process of adapting models for optimal empirical data fit across all methods.
Our results show a more precise model fit than previous studies using the same dataset, underscoring the potential of our methods for enhancing complex pharmacokinetic models, contributing to the evolution of pharmacokinetic modeling and the refinement of therapeutic strategies.

\end{document}